\documentclass[11pt]{article}

\usepackage[T1]{fontenc}
\usepackage[utf8]{inputenc}
\IfFileExists{lmodern.sty}{\usepackage{lmodern}}{}
\usepackage{textcomp}

\usepackage[letterpaper,margin=1in]{geometry}
\usepackage{ragged2e}
\usepackage{indentfirst}
\usepackage{amsmath}
\usepackage{amssymb}
\usepackage{graphicx}
\usepackage{float}
\usepackage[font=small,labelfont=bf,justification=justified,
            singlelinecheck=false]{caption}

\usepackage[numbers,sort&compress]{natbib}

\usepackage[hidelinks,breaklinks]{hyperref}
\hypersetup{
  pdftitle={Energy-Efficient Non-Volatile Photonic Switching via
            Composition-Engineered Sn-Doped GST},
  pdfauthor={X. Zhang, D. Vaz, C.-Y. Kao, M. Tamura, X. Feng,
             C. A. Rios Ocampo, B. J. Shastri, N. Youngblood},
  pdfkeywords={phase change materials; photonic integrated circuits;
               neuromorphic computing; non-volatile photonics;
               low energy switching}
}

\newcommand{\threesubsection}[1]{\par\medskip\noindent\textit{#1}: }

\begin{document}
\justifying

\begin{center}
{\LARGE\bfseries Energy-Efficient Non-Volatile Photonic Switching via Composition-Engineered Sn-Doped GST\par}

\vspace{1.4em}

{\large
Xingyu Zhang$^{1}$,
Daniel Vaz$^{1}$,
Chi-Yi Kao$^{1,2}$,
Marcus Tamura$^{3}$,
Xiong Feng$^{1}$,
Carlos A. R\'ios Ocampo$^{4,5}$,
Bhavin J. Shastri$^{3}$,
Nathan Youngblood$^{1,*}$\par}

\vspace{1.2em}

{\small\itshape
$^{1}$Department of Electrical and Computer Engineering, University of
Pittsburgh, Pittsburgh, Pennsylvania 15261, USA\\[2pt]
$^{2}$Graduate Institute of Electronics Engineering, National Taiwan
University, Taipei, Taiwan\\[2pt]
$^{3}$Centre for Nanophotonics, Department of Physics, Engineering Physics
\& Astronomy, Queen's University, Kingston, Ontario K7L 3N6, Canada\\[2pt]
$^{4}$Department of Materials Science \& Engineering, University of Maryland,
College Park, Maryland 20742, USA\\[2pt]
$^{5}$Institute for Research in Electronics and Applied Physics, University
of Maryland, College Park, Maryland 20742, USA\par}

\vspace{0.9em}

{\small $^{*}$Corresponding author:
\href{mailto:nathan.youngblood@pitt.edu}{nathan.youngblood@pitt.edu}\par}
\end{center}

\vspace{0.5em}

\begin{abstract}
\noindent
Integrated phase-change photonics allows non-volatile optical switching with no static power draw, but the electrical energy needed to reversibly switch the material's state limits practical scaling. Efforts to improve energy efficiency have mostly focused on device optimization, such as redesigning the integrated microheater or waveguide. Here, we take a different, materials-centered approach by alloying Sn to Ge$_2$Sb$_2$Te$_5$ (GST) without altering the device structure or CMOS compatibility. We explore Sn concentrations ranging from 0 at.\% to 20 at.\% on waveguide-integrated PN and PIN heaters and observe a reduction in both the amorphization and crystallization energies in the 5--10 at.\% range. At higher Sn content this trend reverses and the switching energy increases, which is suggestive of phase segregation when combined with TEM compositional analysis. A 10 at.\% Sn device continues to switch over 1000 cycles, with observable degradation in the extinction ratio from 4.5 $\pm$ 0.3 dB to 2.4 $\pm$ 0.3 dB. Post-cycling imaging shows PCM migration and void formation which we believe can be further improved by enhanced encapsulation. Overall, Sn alloying offers a route to lower programming energy in integrated phase-change photonics without negatively affecting the switching speed relative to un-doped GST.
\end{abstract}

\noindent\textbf{Keywords:} \textit{Phase Change Materials, Photonic Integrated Circuits, Neuromorphic Computing, Non-Volatile Photonic, Low Energy Switching}

\vspace{1em}

\section{Introduction}
Phase change materials (PCMs) occupy a singular position in the landscape of functional optical materials. Their ability to reversibly toggle between amorphous and crystalline phases makes them uniquely suited to serve as the active element in non-volatile, reprogrammable devices, which has propelled PCMs from their origins in optical disc storage \cite{Chen2019UnconventionalStorage, Wang2017ApplicationTaxonomy, Ohta2000OverviewTechnology} into an expanding frontier of applications, including electronic and photonic neuromorphic computing \cite{Xu2020RecentMaterials, Zhou2024FabricationComputing, Nandakumar2018AComputing}. For these applications, the ability to store and update synaptic weights in a non-volatile, energy-efficient manner is a central challenge since the programming energies and latencies required to update phase-change weights during the training process or matrix-loading can dominate over computational costs \cite{Yang2023Processing-in-MemoryMemory}.

Within the PCM family, Ge$_2$Sb$_2$Te$_5$ (GST)\cite{Rios2015IntegratedMemory, Wuttig2017Phase-changeApplications, Zheng2018GST-on-siliconPlatform, Adya2025Non-volatileModulators, Adya2024Post-processingProcess} remains the benchmark material for photonic integration due to its high optical contrast ($\Delta n > 2$), fast switching speeds ($<$100 ns), CMOS-compatibility \cite{Khaddam-Aljameh2022HERMES-CoreAADCs}, and high endurance as demonstrated in electronic memory. Low-loss alternatives such as Sb$_2$Se$_3$ \cite{Sun2025MicroheaterPhotonics, Rios2022Ultra-compactMaterials, Blundell2025UltracompactThickness, Wei2023ElectricallyCapability, Delaney2020Asub3/sub, Delaney2021NonvolatileMaterial} and Ge$_2$Sb$_2$Se$_4$Te$_1$ (GSST) \cite{RahimiKari2025High-speedPlatform, Liu2024Non-volatileSwitch, Song2019DesignSwitches} have attracted considerable attention for phase-only non-volatile control, yet their slower crystallization speeds require much longer switching pulses (i.e., several milliseconds) to full recrystallize compared to GST, thus requiring much greater switching energies and higher programming latency. While significant progress has been made in the phase-change photonics community regarding integrated microheater design, cycling endurance, and material integration, the high melting temperature ($\sim$900 K) of GST, Sb$_2$Se$_3$, GSST, and other phase-change chalcogenides, remains a major limitation for large-scale, real-world applications.


Tin (Sn) doping represents a promising strategy for addressing this energy challenge from within the GST material system itself. Unlike nitrogen doping, which aims to improve thermal stability and cyclability \cite{Xia2024SevenRecognition, HoriiARAM, Lai2005Nitrogen-dopedMemory} or copper doping to improve the switching speed \cite{Gao2019EffectFilm}, incorporation of Sn into the GST lattice substitutes weaker Sn-Te bonds ($\sim$359.8 kJ/mol) for stronger Ge-Te bonds ($\geq$396 kJ/mol) \cite{Bai2015EffectLaser, Song2007PhaseMaterial}, significantly reducing the cohesive energy of the network and lowering the activation barrier for the amorphous-to-crystalline reversible transition. This bond-level modification manifests macroscopically as reductions in phase transition temperatures, accelerated crystallization kinetics, and altered grain nucleation and growth dynamics \cite{Bai2015EffectLaser, Yin2019EnhancedEvolution}---all of which directly translate into a reduced energy threshold for reversible switching at the device level. However, the translation of these material-level insights into quantified, device-level metrics on an integrated photonic platform has not been systematically established.

Despite the clear materials rationale for Sn doping, its realization in functional photonic devices has been remarkably limited. Recent work achieving high endurance in programmable silicon photonics employed Sn-GST as the active medium, yet directed its investigation entirely toward the optimization of external ITO heater geometries, which left the compositional degree of freedom of Sn-GST and its influence on the overall switching energy unexplored \cite{Xia2024UltrahighHeater}. A previous study that directly examined Sn doping effects in nanophotonic switching devices was confined to very low concentrations (below 2 at.\%) achieved by ion implantation and have relied exclusively on free-space optical switching of the material \cite{Lazarenko2022LowFilms}. Consequently, no study has established a systematic correlation between Sn-GST composition and the resulting device performance on an electrically driven, CMOS-compatible photonic platform.

This work addresses these gaps through a comprehensive compositional investigation of Sn-doped GST in integrated silicon photonics, spanning pure GST and 5\%, 10\%, and 20\% Sn atomic concentrations on a silicon-on-insulator (SOI) waveguide platform operating in the C-band. A central finding is that switching energy can be substantially reduced through compositional tuning alone, resulting in an up to $\sim$4$\times$ reduction for crystallization and $\sim$1.5$\times$ for amorphization dependent on heater's type. This positions Sn-GST as a potential substitute for GST in applications requiring energy-efficient photonic integration directly compatible with existing foundry processes. By extending the compositional investigation beyond previously studied concentration ranges, this work further uncovers new regimes of Sn-GST behavior and its consequences for switching energy, optical contrast, and device characteristics. Additionally, we explore the switching energies of these different Sn-GST compositions on the same foundry-fabricated chip, allowing us to minimize device-to-device variations across different dies which could obscure compositional-dependent effects. Collectively, these results establish quantitative compositional guidance for Sn-GST in non-volatile photonic integration and provide materials design principles necessary to realize energy-efficient, nonvolatile photonic memory for next-generation computing systems.

\section{Results and Discussion}
\subsection{Illustration of Device Scheme and Concept}
We illustrate our objective in \textbf{Figure 1(a)}, the schematic of intended mechanism, where  incorporating Sn into the GST lattice is expected to reduce the average bond strength and thereby lower the kinetic barrier for the phase transition, reducing the energy threshold for a transmission change. This manifests itself by a reduction in the applied voltage pulse amplitude required to achieve either amorphization or crystallization of the Sn-GST relative to un-doped (pure) GST. To experimentally explore the properties of Sn-doped GST, we designed an array of reconfigurable integrated photonic Mach-Zehnder Interferometers (MZIs), which was subsequently fabricated at a commercial foundry (AMF). Each device consists of a reference arm and a switching arm with embedded microheater (either PN or PIN) as shown in \textbf{Figure 1(b)}. To prevent unwanted interference during switching threshold and endurance measurements, the reference arm window (12 $\mu$m in length) was fully covered by the PCM to effectively suppress its optical transmission after hot plate annealing, thereby eliminating phase-shift contributions that would otherwise complicate real-time monitoring of transmission change. Additionally, this fully covered section serves as characterized regions for compositional mapping via Scanning Electron Microscopy (SEM) and Energy-Dispersive X-ray Spectroscopy (EDS). For the switching arm, we patterned the PCM to be 2 $\mu$m in length, enabling us to measure optical transmission in both states even for compositions with high optical contrast. Electrical switching is achieved through sending a forward-biased pulse through waveguide-integrated PN or PIN microheaters. Thermal simulations for both microheater types are shown in \textbf{Figure 1(c)}. To directly compare the relative switching energies of the different compositions of Sn-GST, we performed all of our experimental measurements on a single chip with identical microheaters (one PN and one PIN microheater for each composition of Sn-GST) to minimize chip-to-chip fabrication variability. Optical micrographs of the MZI array can be seen in \textbf{Figure 1(d)} with detailed dimensions of our oxide window opening to the waveguide shown in the SEM image.
\begin{figure}
    \centering         
    \includegraphics[width=1\textwidth]{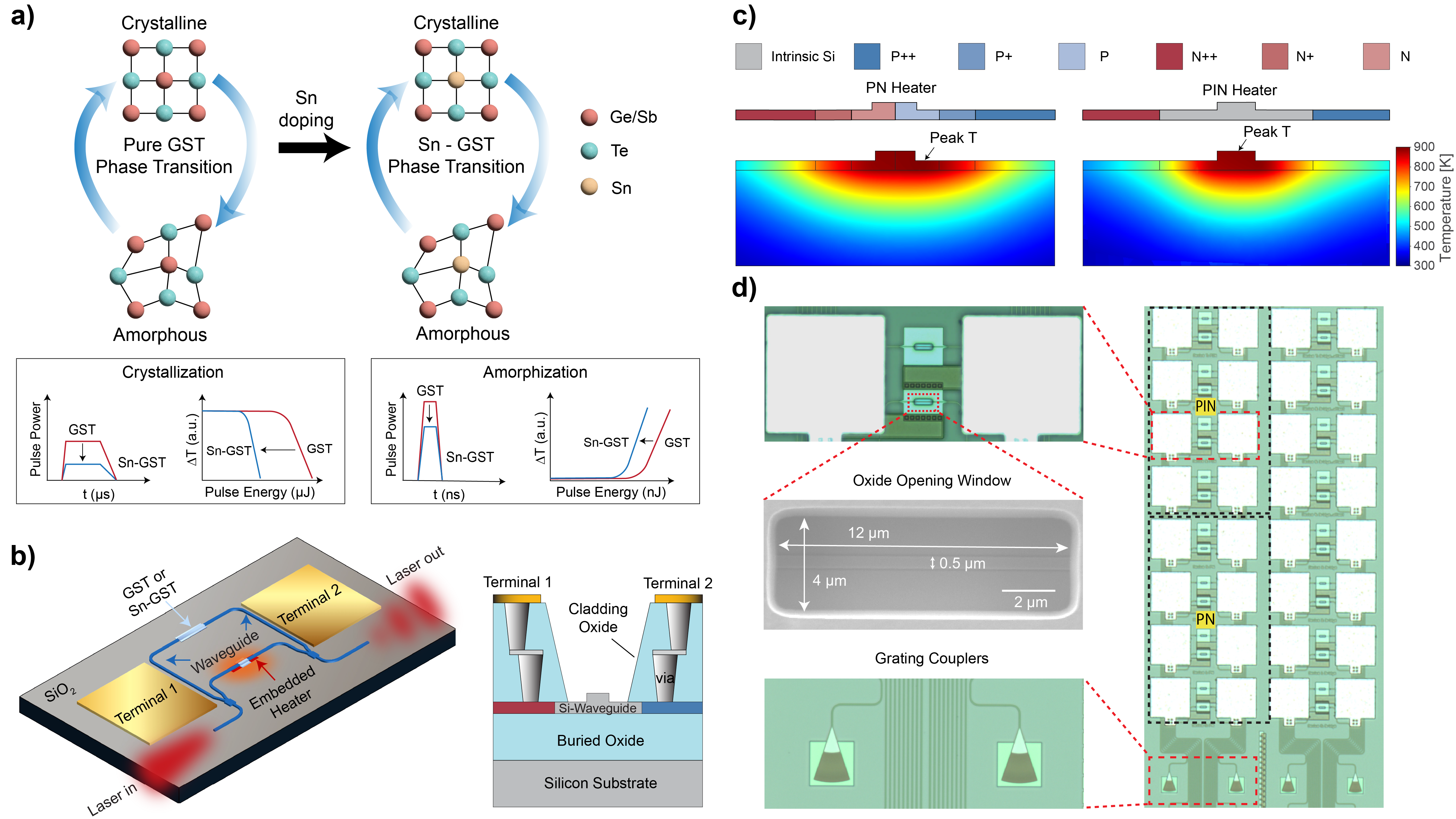} 
    \caption{\textbf{Sn-doped GST for energy-efficient switching.} (a) Conceptual illustration of Sn-doping and its effects on switching energy. Sn substitution into Ge sites reduces the average bond strength of the crystalline lattice, lowering the energy barrier for amorphization relative to undoped GST and reducing the energy threshold for switching. (b) Illustration of the integrated platform used to characterize the switching energies and optical transmission of the PCM. (c) Device cross-section and microheater designs (PN vs PIN) with accompanying thermal simulation of the heat distribution after an amorphization pulse (8.28V/450 ns for PN, 7.02V/450 ns for PIN). (d) Optical and SEM micrographs showing the chip layout and oxide window along with Si waveguide dimensions as well as grating couplers.}
    \label{fig:concept}
\end{figure}
\subsection{EDS Mapping and Compositional Analysis}
Four different batches of devices with distinct doping concentrations were prepared by sputtering deposition. Among them, stoichiometric GST-225 and GST with 20 at.\% Sn-doping were deposited with individual sputtering targets, while devices with $\sim$5\% and $\sim$10\% doping concentrations were deposited with rate controlled co-sputtering with a ratio ($R_{GST}$ : $R_{20\%Sn-GST}$) of 3:1 and 1:1, respectively. Detailed rate calibration can be found in Supplementary Information \textbf{Section S4}. Compositional characterization of the deposited films was performed using EDS elemental mapping and quantitative analysis. Here, \textbf{Figure 2(a)} presents an optical micrograph of the device after deposition where a 2 $\mu$m width of PCM on the switching arm of the MZI was confirmed by SEM. To prevent oxidation during switching, a 60 nm of Al$_2$O$_3$ capping layer was deposited using ALD. The EDS elemental maps, shown in \textbf{Figure 2(b)}, were acquired solely on the reference arm region, revealing spatially uniform distributions of Ge, Sb, Te, and Sn across all compositions investigated, confirming homogeneous co-sputtering without elemental clustering across the film area. The progressive increase in Sn signal intensity observed from pure GST through the intermediate regime to the 20 at.\% Sn-GST corroborates the compositional tunability of Sn via co-sputtering rate control. The quantitative EDS analysis (\textbf{Figures 2(c)-(d)}) further substantiates these findings as the Sn L$\alpha$ emission peak at 3.44 keV exhibits monotonic growth with increasing Sn content. The extracted atomic compositions confirm that pure stoichiometric GST contains a negligible Sn content of 0.7 at.\%, in line with baseline noise, whereas the rate ratios of 3:1, 1:1 and 20 at.\% Sn-GST samples yield Sn concentration of 5.3, 10.9, and 20.2 at.\%, respectively. Collectively, these results demonstrate reproducible control of Sn content through co-sputtering rate control. We note, however, that the resulting series is not just a pure Sn-for-Ge substitution and is more accurately described as a Te-deficient quaternary Ge--Sb--Sn--Te alloy than as Sn-doped GST-225. Because Ge/Sb/Te stoichiometry independently affects crystallization kinetics, melting behavior, optical constants and phase stability, the trends reported below reflect the combined effect of Sn incorporation and Te depletion and cannot be uniquely attributed to Sn alone.
\begin{figure}
    \centering
    \includegraphics[width=0.8\textwidth]{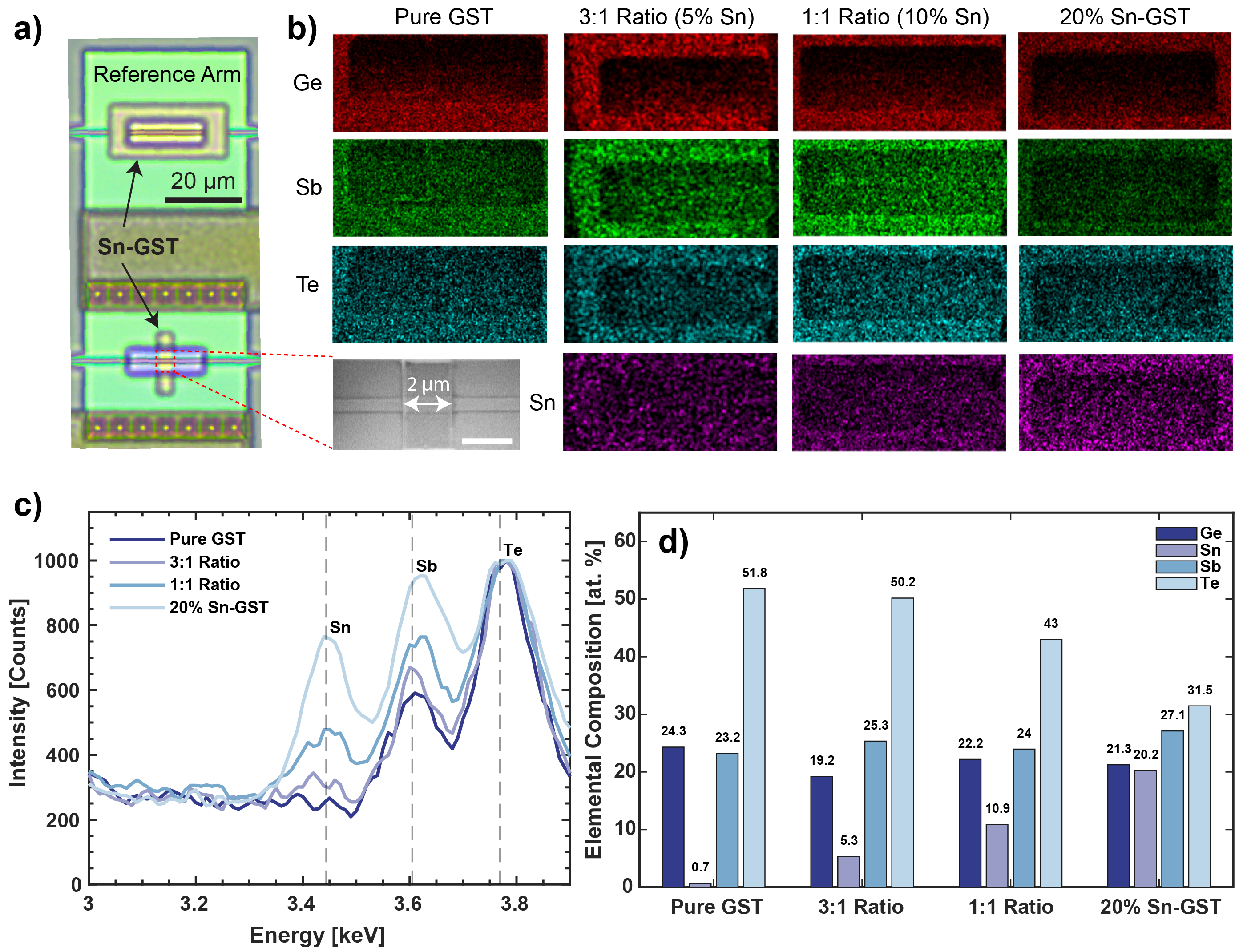} 
    \caption{\textbf{Device architecture and compositional characterization of co-sputtered Sn-GST thin films}. (a) Optical micrograph of the MZI device after both arms deposited with GST or Sn-GST. Inset shows the width of material deposited on the Si waveguide (2 $\mu$m scale bar). (b) EDS elemental maps acquired for pure GST, co-sputtered films at rate ratios $R_{GST}$ : $R_{20\%Sn-GST}$ = 3:1 and 1:1, and 20 at.\% Sn-GST, confirming spatially homogeneous elemental distributions across all compositions. (c) EDS spectra showing a systematic growth in Sn L$\alpha$ peak intensity with increasing Sn content. (d) Quantitative EDS elemental composition analysis in at.\% across all 4 film compositions, demonstrating precise compositional control through co-sputtering rate modulation. }
    \label{fig:eds}
\end{figure}
\subsection{Sn-Dependent Complex Refractive Index}
The complex refractive index $\tilde{n}=n+ik$ of pure and Sn-doped GST was determined by spectroscopic ellipsometry for as-deposited films and on the same films after hotplate annealing at $200^\circ$C (\textbf{Figure 3(a)-(b)}). The amorphous state refractive index was found to range from $4.5-5.3$ with increasing Sn content, with relatively minor differences in the refractive index below 20\% Sn. After crystallization, we see a modest decrease in the refractive index in the near-IR as we increase Sn content from pure GST to 10 at.\% Sn GST. For high doping concentrations of 20 at.\% Sn, however, the refractive index contrast is significantly reduced. These trends are consistent for the extinction coefficient as well, with Sn doping significantly reducing the contrast in $\Delta k$ above 10 at.\% Sn, while slightly raising the loss in the amorphous state below this threshold.

Since we are targeting amplitude modulation in the C-band, the relevant figure of merit (FOM) becomes the relative absorption modulation depth $ {\Delta k}/{k_{am}}$ as discussed in \cite{Youngblood2023IntegratedMemristorsb}. As shown in \textbf{Figure 3(c)}, the FOM falls monotonically from 9.0 for pure GST to 0.3 for 20 at.\% Sn doping due to the increasing absorption in the amorphous state which introduces unwanted loss. The absolute contrast $\Delta k$ is largely preserved up to 10 at.\% (ranging from $1.4-1.55$) but collapses at 20 at.\%. The index contrast $\Delta n$ likewise decreases from $>$2.5 for Sn doping at or below 10 at.\%, down to 0.63 for 20 at.\% Sn doping. The results suggest a clear trade-off for photonic design. Although Sn doping is known to lower the switching energy through weaker Sn-Te bonding, it does so at the expense of increased loss in the C-band. The optical contrast is retained while the loss remains relatively low for Sn $\leq$10 at.\%, but the advantages of Sn doping is essentially lost at 20 at.\%, suggesting $\sim$10 at.\% Sn as the practical upper bound for high-contrast C-band amplitude modulation. 
\begin{figure}
    \centering
    \includegraphics[width=1\textwidth]{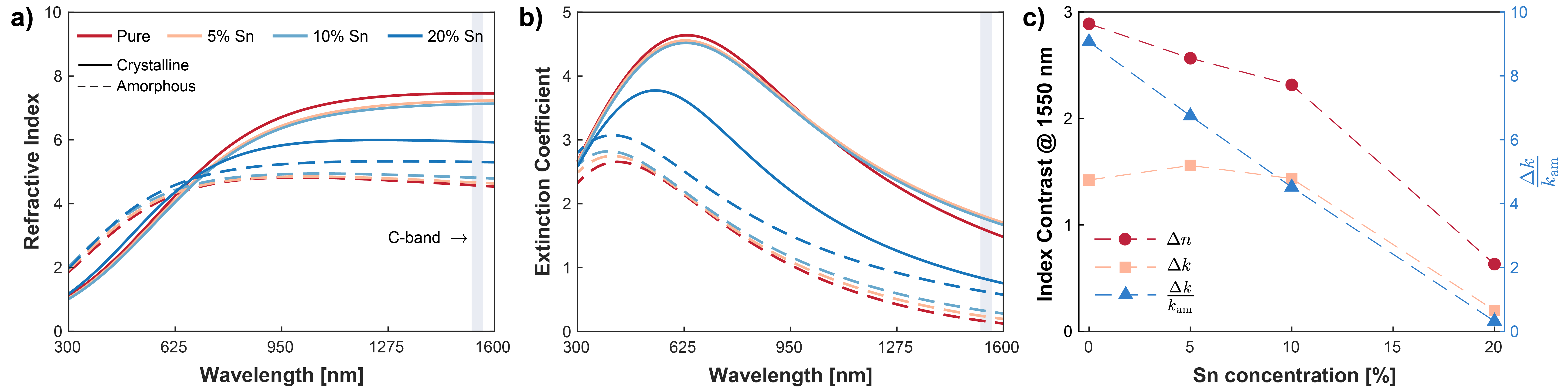}
    \caption{\textbf{Complex refractive index of Sn-doped GST measured via spectroscopic ellipsometry.} (a) Refractive index n and (b) extinction coefficient k as a function of wavelength for pure GST and thin films with 5 at.\%, 10 at.\% and 20 at.\% Sn. Solid and dashed lines denote the crystalline and amorphous phases, respectively. The shaded band marks the telecom C-band ($1530-1565$ nm). (c) Optical contrast for n and k as well as the relative absorption modulation ${\Delta k}/{k_{am}}$ at 1550 nm as a function of Sn concentration, highlighting the design trade-offs for photonic devices using Sn-doped GST.}
    \label{fig:ellipsometry}
\end{figure}
\subsection{Energy Threshold Tuning with Sn Dopant Concentration}
Having successfully characterized the composition and optical properties of Sn-GST, we proceed to characterize the switching energy threshold needed for the onset of amorphization and crystallization. For this purpose, we define the normalized transmission change ($\Delta T$) as:
\begin{equation}
    \Delta T = \frac{T-T_{cry}}{T_{max}}
\end{equation}
where $T$ denotes the stabilized transmission recorded after each applied pulse, $T_{cry}$ is the crystalline transmission state after annealing, and $T_{max}$ corresponds to the maximum transmission observed for each individual device, representing complete switching to the amorphous state. This normalization accounts for inherent device-to-device variation in absolute transmission levels and switching contrast, enabling consistent quantitative comparison across all material compositions.
Pulse energy was determined from oscilloscope-recorded waveforms through time-domain integration:
\begin{equation}
    E=\int P(t) dt = \int V_{device}(t)\times I(t) dt=\int V_{device}(t)\times \frac{V_{sense}(t)}{R_{sense}}dt
\end{equation}

For determining the switching energy of each pulse, we placed a calibrated sense resistor $R_{sense}$ in series with the device to enable accurate real-time current extraction. To measure the voltage drop across $R_{sense}$, the voltages immediately before and after the sense resistor were recorded and their difference yielded $V_{sense}(t)$ from which the time-dependent current is derived. For amorphization, we used 450 ns pulses with a 7 ns falling edge and variable amplitude, while for crystallization we employed 80 $\mu s$ pulses with a 30 $\mu s$ falling edge. The full details of the experimental setup are provided in Supplementary Information \textbf{Section S2}, while in \textbf{Section S3} we characterize the pulse energy as a function of pulse amplitude for both PN and PIN microheaters. Both heater types exhibit broadly similar energy-amplitude scaling, with their outputs converging at high voltage amplitudes. However, at lower applied voltages, the PN heater consistently delivers less energy than the PIN heater under identical pulse conditions. This discrepancy is attributed to slight differences in the low-voltage resistance characteristics of the two junctions. Given the extended pulse duration required for crystallization, even modest per-pulse energy differences can result in notable disparities in temperature experienced by the PCM.


\textbf{Figures 4(a)} and \textbf{4(c)} reveal that the amorphization threshold energy (i.e., the minimum energy required to observe an increase in transmission) almost unilaterally decreases with increasing Sn concentration for both heater types. The estimated switching pulse energies threshold from pure GST to 20 at.\% Sn doped are approximately 303 nJ, 216 nJ, 185 nJ and 198 nJ, respectively for the PN microheater, and 203 nJ, 190 nJ, 172 nJ, and 156 nJ for PIN. We attribute the small increase in energy at 20 at.\% for the PN microheater to be caused by the limited resolution of our pulse generator at high voltages which has coarser control over the pulse-to-pulse energy than all-optical switching methods. With the reported difference in melting temperature between 10 at.\% and 20 at.\% doped GST just $9^{\circ}$C apart \cite{Song2007PhaseMaterial}, it is quite challenging to resolve this difference with our electrical pulses. Regardless, our results show a consistent trend toward decreased amorphization energy with increasing Sn concentration. This is consistent with a reduction in the cohesive energy of the network. Across all compositions, the PN heater requires moderately higher input energy to trigger amorphization compared to the PIN heater, with the difference being most pronounced for pure GST at approximately 100 nJ. We attribute this to the different thermal profiles generated by the two junction architectures. In the PIN microheater, the intrinsic layer separating the p-and n-doped regions produces a more spatially uniform heat distribution than the PN microheater, with the peak temperature of the heater centered in the intrinsic region \cite{Zheng2020NonvolatileHeater,Erickson2022DesigningMicroheaters}. According to our PN microheater simulations, there is an asymmetric thermal profile concentrated toward the p-doped side (see \textbf{Figure 1(c)} and Supplementary \textbf{Section S10}) with a more pronounced thermal gradient. Since a successful amorphization requires a sufficient volume fraction of material be simultaneously melted and rapidly quenched, the spatial mismatch between the localized heat source and the full active volume in the PN configuration necessitates higher total input energy to achieve the critical melt threshold.

\begin{figure}
    \centering
    \includegraphics[width=0.9\textwidth]{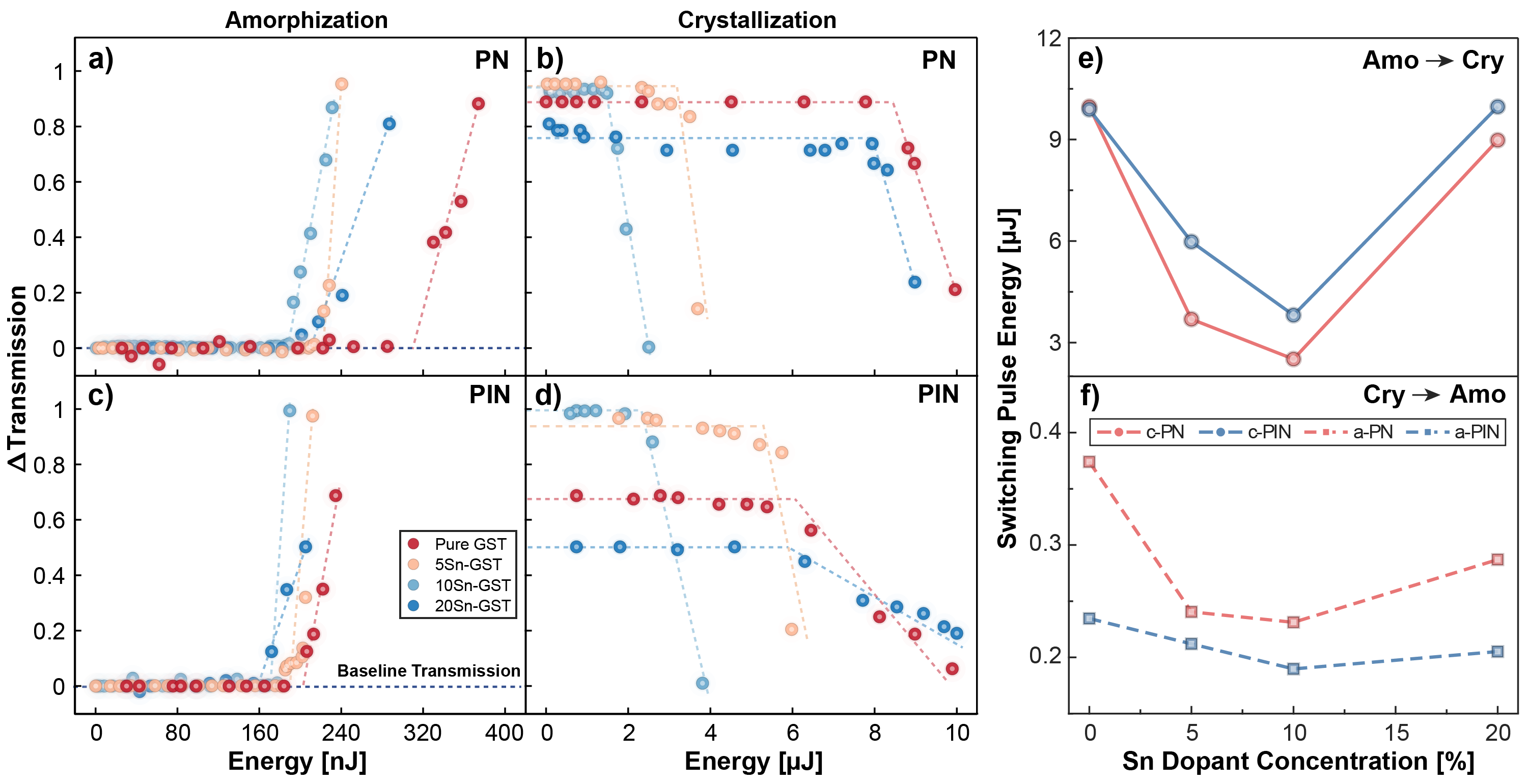}
    \caption{\textbf{Pulse energy required for electrical switching of Sn-doped GST integrated on PN and PIN microheaters.} (a)-(d) Change in normalized optical transmission ($\Delta T$) vs single-pulse energy for all four doping conditions, actuated by (a)-(b) PN and (c)-(d) PIN microheaters for amorphization and crystallization. (e)-(f) Switching pulse energy required for full (e) crystallization and (f) amorphization for all four doping concentrations. All switching curves show a minimum at 10\% Sn, suggesting an optimal window for Sn dopant concentration. }
    \label{fig:energy}
\end{figure}

The crystallization behavior exhibits a distinctly non-monotonic dependence on Sn concentration, as shown in \textbf{Figure 4(b)} and \textbf{(d)}. Devices incorporating 5 at.\% and 10 at.\% Sn display substantially reduced crystallization threshold energies relative to pure GST, while the 20 at.\% Sn sample shows a notable increase in switching threshold relative to 10\% Sn, with the threshold energy approaching that of pure GST. This  is possibly due to the formation of a SnTe secondary phase upon exceeding the Sn solubility limit in the GST matrix, which raises the effective crystallization temperature and counteracts the nucleation-promoting benefit of Sn incorporation. This precipitation of a secondary phase in a Sn-alloy system is not confined to Sn-GST alone. A study has shown in the Sn-Zn-Sb alloy system, a secondary phase like SnSb can precipitate out along with ZnSnSb$_2$ during heat treatment over 11.1 at.\% Sn doping \cite{Zhang2019ImprovedApplications}. For the 5 at.\% and 10 at.\% Sn compositions, the PN heater yields lower crystallization threshold energies of 3.22 $\mu$J and 1.46 $\mu$J, respectively compared to the threshold energies of the PIN heater at 5.31 $\mu$J and 2.44 $\mu$J. This behavior can be understood from the nucleation-dominated crystallization kinetics of GST. In this process, crystalline nuclei first form locally, then grow and coalesce into a continuous crystalline network. Existing studies have shown Sn doping lowers the nucleation energy barrier in GST, facilitating the onset of nucleation at reduced thermal input \cite{Bai2015EffectLaser, Li2019CrystallizationLaser}. Consequently, the locally concentrated heating profile of the PN junction, while disadvantageous for the volume-melting requirement of amorphization, proves sufficient heat to trigger local nucleation, sustain subsequent grain growth, and coalescence at lower total pulse energy. The spatially uniform heating of the PIN heater, by contrast, distributes thermal energy across a larger volume, which is less efficient to initiate nucleation in these Sn-doped compositions. 


The pulse energies required to fully amorphize and recrystallize the PCM across all compositions are compiled in \textbf{Figure 4 (e)-(f)}. The resulting dependence on Sn concentration identifies 10 at.\% as the optimal doping level among our four concentrations, minimizing the switching energy for both amorphization and crystallization. This non-monotonic behavior delineates a practical compositional window for Sn-GST in integrated photonic devices. To date, it is debated how Sn doping affects the phase transition process in GST. It is widely agreed that the decrease in cohesive energy lowers the melting temperature and consequently the amorphization (reset) energy---a trend that is largely consistent across literature. For crystallization, however, the picture is far less clear. Some studies report that a low dopant concentration lowers the crystallization temperature while a high concentration raises it \cite{CharacteristicsMemory}, while others report the opposite trend \cite{Yin2017TheSn}. Still others find an almost monotonic dependence with Sn doping \cite{Li2019CrystallizationLaser}. With this context, we believe our findings provide valuable results for the GST system and phase change material community.

\subsection{Material Characterization and Phase Segregation}
We performed X-ray diffraction (XRD) on all four Sn-GST compositions as shown in \textbf{Figure 5}. The diffraction patterns in \textbf{Figure 5(a)} confirm that all compositions present metastable face-centered-cubic (fcc, rock-salt) phase of GST as evidenced by the characteristic (111), (200), (220), and (222) peaks exhibited across all samples, which shows good agreement with existing literature. With increasing Sn content, two noticeable changes are clearly observed. First, the diffraction peaks shift progressively toward lower $2\theta$ angles, indicating an expansion of the lattice parameter. This expansion is consistent with the substitution of Ge (1.22 \AA) by the larger Sn (1.40 \AA) atom on the cation sublattice, which increases the average interplanar spacing in accordance with Bragg's law and lowers the diffraction angles. Second, the diffraction peaks broaden at higher Sn concentrations, which may reflect a reduction in crystallite size and/or an increase in micro-strain induced by substitution. Most notably, the 20 at.\% Sn-GST sample exhibits a set of additional reflections that can be indexed to the SnTe (200), (220), and (222) planes, providing direct evidence for the formation of a crystalline SnTe secondary phase once the Sn content exceeds the solubility limit within the GST matrix.
\begin{figure}
    \centering
     \includegraphics[width=1\textwidth]{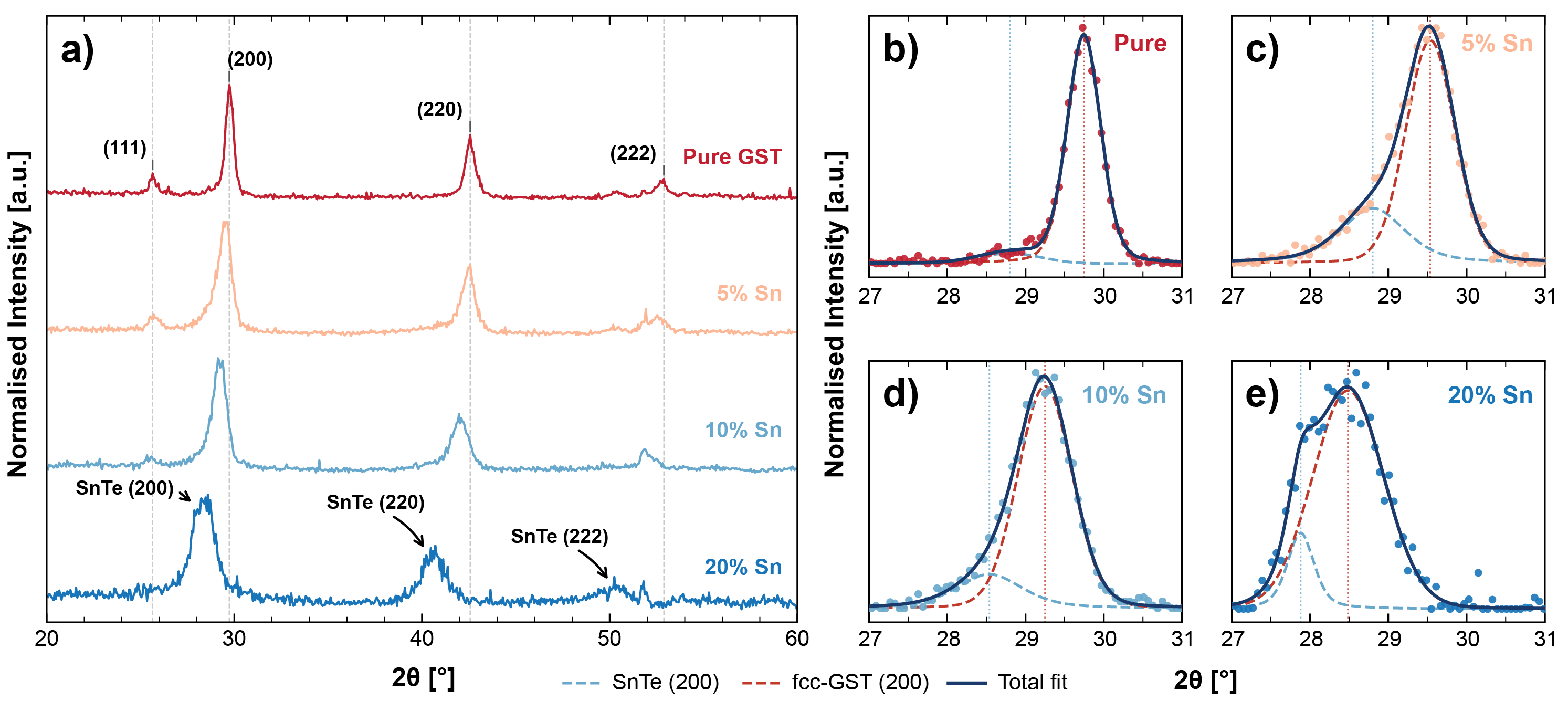}
    \caption{\textbf{Structural characterization and phase evolution of GST and Sn-doped GST films.} (a) XRD patterns of pure GST, 5\% Sn-GST, 10\% Sn-GST and 20\% Sn-GST. All compositions crystallize in the fcc/rock-salt phase, indexed by the (111), (200), (220), and (222) reflections. The 20\% Sn-GST shows additional SnTe (200), (220), and (222) reflections pointed by arrows indicating a SnTe secondary phase segregation. (b)-(e) deconvolution of the (200) reflection for (b) pure GST, (c) 5\% Sn-GST, (d) 10\% Sn-GST, and (e) 20\% Sn-GST. Each experimentally measured peak is fitted with two pseudo-Voigt components, fcc-GST (200) and SnTe (200). The most pronounced double peak feature is observed at 20\% Sn-GST film.}
    \label{fig:xrd}
\end{figure}


To resolve the phase evolution in greater detail, the dominant (200) reflection was analyzed for each composition over the  $27-31^\circ$ range, as presented in \textbf{Figures 5(b)-(e)}. Each measured profile was fit using two pseudo-Voigt functions, corresponding to the fcc-GST (200) and SnTe (200) reflections, with the summed profile (total fit) reproducing the experimental data across all compositions. For pure GST curves in \textbf{Figure 5(b)}, the reflection is well described by a single fcc-GST component, with a negligible SnTe contribution. As the dopant concentration increases (\textbf{Figure 5(c)-(d)}), the fcc-GST component shifts progressively to lower angles, again reflecting Sn-induced lattice expansion, while the SnTe component is still relatively weak.  Only at 20 at.\% Sn, shown in \textbf{Figure 5(e)}, does the low-angle component become pronounced enough to produce a clear double-peak profile, indicating phase segregation at this composition.

To confirm the phase segregation at 20 at.\% Sn, cross-sectional TEM was carried out to validate the XRD results. We examined two samples: a film annealed at $200^\circ$C which we used for ellipsometry and an as-deposited film with no thermal treatment. The as-deposited sample was capped with $\sim$60 nm of Al$_2$O$_3$, the same passivation used in our actual devices, to prevent ambient oxidation during cycling. The annealed specimen was deliberately left uncapped to simplify the optical model used for the ellipsometry fit. The two samples came from separate runs but followed an identical deposition recipe, so we expect their initial states to be equivalent and run-to-run differences to be negligible. 
\begin{figure}
    \centering
     \includegraphics[width=1\textwidth]{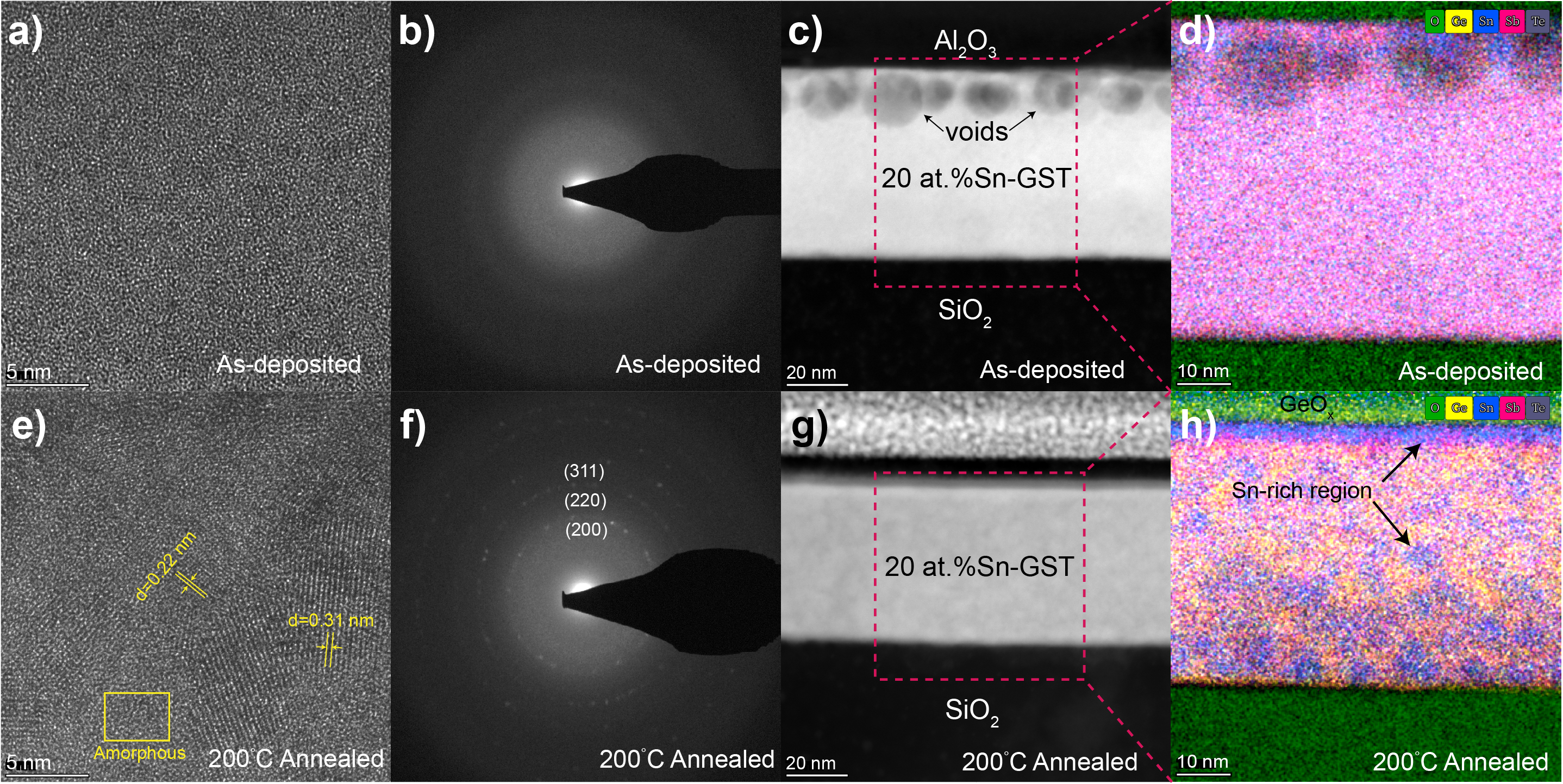}
    \caption{\textbf{TEM characterization of phase segregation in 20 at.\% Sn-doped GST films.} (a)-(d) represents the TEM analysis on 20\% Sn-GST for as-deposited with (a) HRTEM, (b) SAED, (c) HAADF and (d) cross-sectional EDS mapping in at.\%. (e)-(h) TEM analysis on thermally annealed (crystalline state) 20\% Sn-GST with the same analysis for comparison.}
    \label{fig:tem}
\end{figure}

The as-deposited film is fully amorphous as confirmed in its HRTEM image (\textbf{Figure 6(a)}) which shows a disordered atomic arrangement with no resolvable lattice fringes, while the SAED pattern (\textbf{Figure 6(b)}) contains only a diffuse halo. Annealing at $200^\circ$C produces observable crystalline features. In \textbf{Figure 6(e)}, we resolve two sets of lattice fringes with spacing of 0.31 nm and 0.22 nm, matching the (200) and (220) planes of a rock-salt phase in GST matrix system, while amorphous regions remain visible across the image, indicating only partial crystallization. The corresponding SAED pattern (\textbf{Figure 6(f)}) confirms this, showing three identifiable reflections indexed to (200), (220) and (311) superimposed on a weak, diffuse background. The SnTe and FCC-GST matrix differ in lattice parameter by only about 5\%, so their interplanar spacings differ by a similar amount, well within the measurement uncertainty of HRTEM fringe spacings and SAED rings. Neither technique can therefore separate the SnTe phase from the GST matrix on its own, so we turned to compositional mapping.

We recorded HAADF-STEM images of both specimens (as-deposited in \textbf{Figure 6(c)} and annealed in \textbf{Figure 6(g)}) and mapped the composition by EDS over the PCM layer. The composite maps (at.\%) are given in \textbf{Figure 6(d)} and \textbf{Figure 6(h)}, and the element by element maps in the Supplementary Information \textbf{Section S13}. The as-deposited film is compositionally homogeneous, whereas the annealed film is clearly segregated throughout the layer, implying segregation develops during initial crystallization. The map of annealed Sn-GST shows two notable features. At the top surface, a $GeO_x$ layer has formed, with a Sn- and Te-rich band directly underneath it. Because this specimen was uncapped, we attribute this near-surface feature to oxidation rather than crystallization where Ge is drawn out of the matrix to form the oxide, leaving Sn and Te behind. Deeper in the film, we see larger clusters of differing composition and we assign these to crystallization-driven segregation. Combined with the HRTEM partial crystallization results, our analysis indicates that phase segregation during crystallization can impede crystallization kinetics, thereby increasing the energy barrier. We also observe voids at the Sn-GST/Al$_2$O$_3$ interface in the as-deposited sample, whose origins are unclear. Resolving whether they arise from growth stress, densification, or an interfacial reaction would require a dedicated study aimed at improving film quality for photonic PCM devices, which is beyond the scope of this paper. However, it is worth noting these voids are not present in the un-capped film after annealing.

Taken together, these results are fully consistent with the pulse switching behavior described above. Low to moderate Sn contents are accommodated within a single rock-salt lattice, which improves the switching kinetics and device performance. Higher Sn concentrations lead to the precipitation of a distinct crystalline SnTe phase which suppresses the crystallization kinetics by raising the effective crystallization temperature and accounts for the observed increase in switching energy at 20 at.\% Sn.

\subsection{Cyclic Performance of 10\% Sn-GST}
Reversible switching behaviors have been demonstrated for all compositions and devices and are included in the Supplementary Information (\textbf{Section S9}). Based on our previous observations of PN microheater performance with low or moderate Sn-doped devices, we selected a 10 at.\% Sn-doped sample for cyclic testing. To localize switching to the waveguide region with greatest overlap with the optical mode, we designed a 5 $\mu$m pattern confined to the Si waveguide on the switching arm, as shown in \textbf{Figure 7(a)}. 


\begin{figure}
    \centering
    \includegraphics[width=1\textwidth]{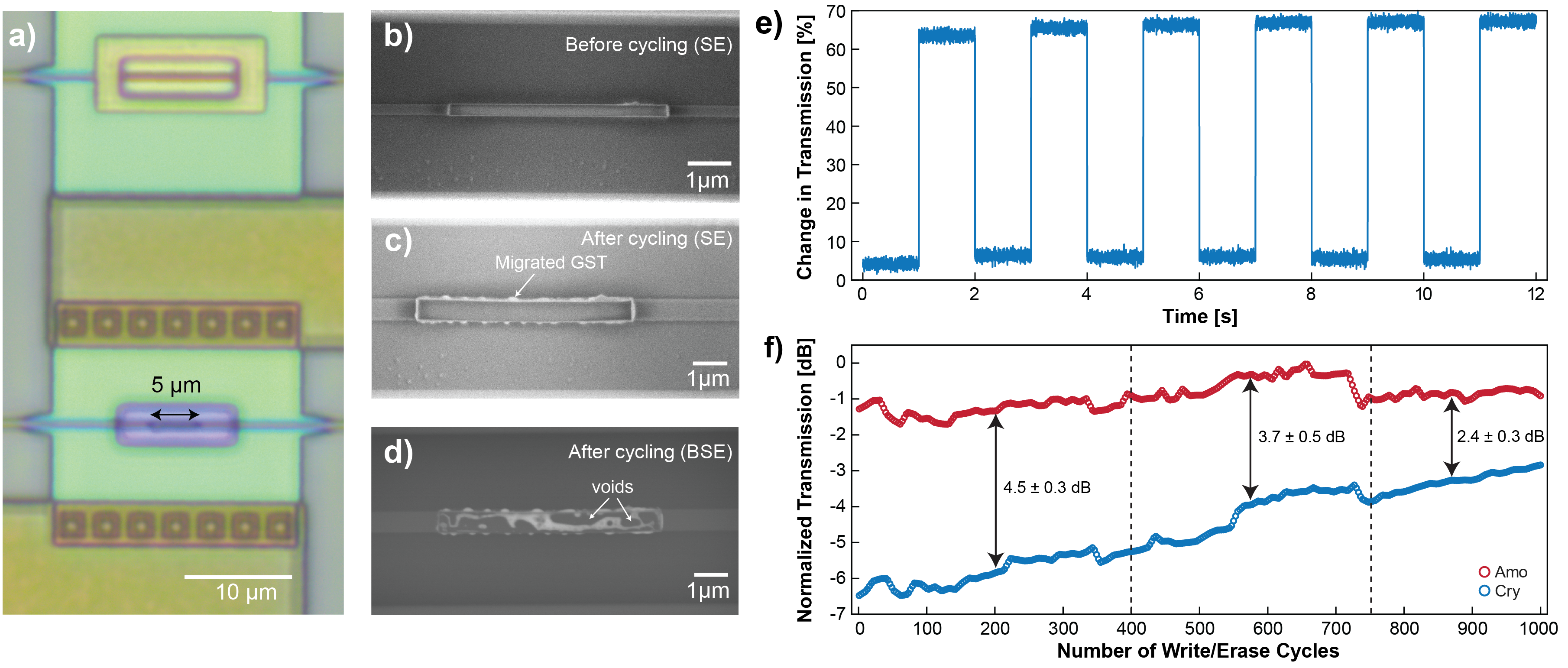}
    \caption{\textbf{Cyclic performance of 10\% Sn-GST.} (a) Optical micrograph showing a Sn-GST patch on top of a Si waveguide with integrated PN microheater. (b)-(d) SEM images of the Sn-GST patch before and after cycling. (b) Secondary electron (SE) image of the as-fabricated device before cycling. (c) SE image of the device after cycling showing PCM migration toward the patch periphery and no ablation. (d) Backscattered electron (BSE) image of device after cycling, enabling imaging of the Sn-GST under the Al$_2$O$_3$ capping layer which reveals void formation from movement of the material. (e) Real-time optical transmission over 6 cycles under alternating between amorphization (8.28V/450 ns) and crystallization (3.12V/80 $\mu$s) pulses. (f) Cycling endurance over 1000 switching cycles. Dashed lines divide the data into 3 regions, showing a progressive reduction attributed to PCM migration and void formation shown in (c) and (d). }
    \label{fig:cycling}
\end{figure}

To clearly demonstrate the binary state switching cyclability of our device, we recorded the relative change in transmission for the device during a series of cyclic amorphization and crystalline pulses. Temporal switching results are shown in \textbf{Figure 7(e)} for the initial 6 cycles, which results in approximately 4.6 dB of Extinction Ratio (ER) for a $\sim$15 nm thin film of 10 at.\% Sn-doped GST (0.92 dB/$\mu$m switching contrast). The transmission states of amorphous (Amo) and crystalline (Cry) Sn-GST over 1000 switching cycles are tracked and presented in \textbf{Figure 7(f)}. The cycling data shows that in the first 400 cycles, the ER is relatively stable, producing an ER of about 4.5 $\pm$ 0.3 dB on average, which is approximately the lifetime applying a failure criterion of a 1 dB loss of extinction ratio from the initial value. After further cycling, degradation in the ER was observed from 400 to 750 cycles, where the average dropped to 3.7 $\pm$ 0.5 dB, which corresponds to 50\% contrast loss criterion. The device continues to switch measurably to 1000 cycles, but with an extinction ratio of 2.4 $\pm$ 0.3 dB, and we describe this as residual rather than stable operation.

To understand the degradation mechanism, we used SEM imaging to compare the device before and after cycling (\textbf{Figure 7(b)-(d)}). In \textbf{Figure 7(b)-(c)}, the images were collected using the secondary electron (SE) detector which is sensitive to the surface conductivity of the material. It is clear from \textbf{Figure 7(b)} that the PCM pattern is well defined and centered on the waveguide. Both \textbf{Figure 7(c)-(d)} are collected after the cyclic measurement shown in \textbf{Figure 7(f)}. In \textbf{Figure 7(c)}, we observed no noticeable damage to the top Al$_2$O$_3$ capping layer, suggesting no ablation or mechanical failure of the PCM structure. However, we noticed material emerging from the edges of the Sn-GST patch which we attribute to material reflowing and migrating away from the center during cycling. To better understand this effect, we applied a higher (15 kV) accelerating voltage and imaged the structure using backscattered electron (BSE) imaging. This mode is sensitive to the atomic properties of the material, giving contrast based on the atomic number (Z) rather than the surface conductivity. Here, we clearly spot the formation of voids and large portions of the PCM reflowing under the cladding layer. Given the Al$_2$O$_3$ passivation layer is patterned during lift-off and there is no coverage on the side of PCM, the observed degradation of ER and material leaving around the edges of the Sn-GST patch make sense. The migration of Sn-GST throughout the cycling measurements changes the optical mode overlap and contributes to variability in the ER. The cyclic endurance can be improved with better PCM patterning, such as confined circular disks \cite{Wu2019Low-LossMaterial} or tapered PCM islands \cite{Dutta2026IncreasedMaterials} surrounded by thick ALD-grown capping layers to prevent the reflow of material and oxidation from the edges \cite{Chen2022BroadbandPhotonics, Golovchak2015OxygenMatrix,Popescu2025UnderstandingTe}.Thus, we expect that proper patterning and passivation techniques would dramatically improve the cycling endurance of Sn-doped GST.

However, We cannot establish that the degradation is fully extrinsic to the material. Melt viscosity, wetting and adhesion to the waveguide, interfacial reactivity and the temperature excursion required per pulse are all composition-dependent, and the reduced cohesive energy that lowers the switching energy of Sn-alloyed GST may equally promote reflow and mass transport at the melt temperature. Distinguish these contributions would require further studies on the degradation mechanism.

\section{Conclusion}
In conclusion, we have demonstrated that Sn doping lowers the cohesive energy of the GST lattice, leading to a significant reduction in the switching energies required for amorphization  and crystallization with a minor increase in optical loss below 10 at.\% Sn. Our results provide detailed characterization of the optical and material properties of Sn-doped GST across a much wider doping range than previously explored. These results have given us insight into the practical limitations of Sn-doping beyond 10 at.\% which leads to phase segregation, resulting in poor optical contrast and higher switching energies. We attributed this phase segregation arose from secondary phase SnTe precipitation, which we further confirmed via XRD and cross-sectional TEM. A representative 10 at.\% Sn-doped GST memory device was continuously switched up to 1000 cycles, with loss in optical contrast traced to PCM migration and void formation rather than mechanical failure or phase segregation, indicating endurance is readily improvable through optimized encapsulation and PCM patterning though more detailed study is still required to fully understand the degradation mechanism. This work provides a promising method to reduce switching energy in phase-change photonics through materials level design, offering a promising route toward energy-efficient photonic memory for reconfigurable photonics and neuromorphic photonic computing.

\section{Experimental Section}
\threesubsection{\textbf{Device Fabrication}}
The devices were fabricated using the active silicon photonics platform (AMFSiP) at Advanced Micro Foundry (AMF). The details of PCM integration through post-processing are included in the Supplementary Information \textbf{Section S1}. With the waveguides exposed through open oxide windows provided by the foundry process, subsequent EBL lithography (Raith EBPG 5150) with PMMA (950 PMMA A4) was performed to define PCM patches on the waveguides (dose used for lithography was 1800 $\mu$C/cm$^2$). After development with MIBK/IPA 1:3 for 1 min, the pure GST and Sn-GST target concentrations were deposited with pure GST and 20 at.\% Sn-GST targets via co-sputtering using Angstrom Engineering Sputtering System with rate control on each target. Details on the rate calibration and co-sputtering process can be found in Supplementary Information \textbf{Sections S4} and \textbf{S5}. Finally, the structure was encapsulated with $\sim$60 nm of protective oxide cladding layer (Al$_2$O$_3$) via atomic layer deposition (Ultratech Fiji G2 plasma) at $100^\circ$C. Lift-off was then performed by leaving the samples in acetone or PG Remover overnight at room temperature.

\threesubsection{\textbf{Experimental Setup and Device Measurement}}
The experimental setup for device testing is presented in Supplementary Information \textbf{Section S2}. For optical transmission measurements, a tunable laser source (Santec TSL-550) was swept across a wavelength range of 1500 nm to 1630 nm. Laser light was delivered to the chip through a 16-channel fiber array, and an in-line fiber polarization controller was tuned to preferentially excite the fundamental quasi-transverse-electric (quasi-TE) mode while maximizing the fiber-to-chip coupling efficiency. The transmitted light was subsequently converted into an electrical signal by a low-noise photodetector (Newport 2011-FC), which was sampled and recorded through a data acquisition (DAQ) board (National Instruments BNC-2110).

For electrical measurements, voltage pulses were synthesized by an arbitrary waveform generator (Rigol DG4102) and boosted by a high-speed voltage amplifier with a fixed gain of 5V/V. The amplified pulses were routed to the on-chip contact pads via a dual-tip ground-signal RF probe (FormFactor FPC-GSG-250). To accurately quantify the electrical energy supplied during each programming event, a sense resistor of 9 $\Omega$ was placed in series with the device, and the voltage difference across its two terminals was read out using a high-frequency oscilloscope (Rigol MSO8204). 

\threesubsection{\textbf{Material Characterization}}
The X-ray Diffraction for each sample was taken from reference chips which were sputtered simultaneously with the devices under test for each Sn-GST concentration. Thin films were characterized by grazing-incidence x-ray diffraction (GIXRD) using a Malvern PANalytical EMPYREAN diffractometer equipped with a PIXcel detector. Cu K$\alpha $ radiation ($\lambda = 1.5406$~\AA) was generated at a tube voltage of 45kV and a current of 40 mA, with a fixed divergence slit of $0.03125^\circ$. To enhance the diffraction signal from the thin film while supressing the substrate contribution, the incident beam was fixed at a grazing angle of $\omega = 0.7^\circ$. Diffraction patterns were collected over a $2\theta$ range of $20^\circ$ to $60^\circ$ with a step size of $0.06^\circ$ and a counting time of 5.28 s per step.

Scanning electron microscopy (SEM) of the four devices was performed using a ZEISS Sigma 500 VP microscope. Secondary electron (SE) images were acquired at an accelerating voltage of 2 or 3 kV, whereas backscattered electron (BSE) images were obtained at 15 kV. Energy-dispersive X-ray spectroscopy (EDS) analysis of all reference arms and targeted devices (Supplementary Information \textbf{Section S10}) was carried out at 15kV as well.

The complex refractive index ($n,k$) of the amorphous and crystalline phases across all four films was determined by spectroscopic ellipsometry over the wavelength range of 300-1620 nm using a Horiba UVISEL Plus Spectroscopic Ellipsometer using the reference chips. The measured spectra were analyzed with a Si/SiO$_2$/GST(dispersion)/surface-roughness optical model, in which the dielectric function of the PCM layer was represented by a Tauc-Lorentz oscillator. 

\threesubsection{\textbf{Device Modeling and Simulation}}
The transient temperature distribution in the PN and PIN microheaters was obtained from a fully coupled electro-thermal finite-element model implemented in Python (FEMWELL, available at \url{https://helgegehring.github.io/femwell/}), using device dimensions identical to the fabricated heaters. The model solves the current-continuity equation $\nabla\cdot(\sigma (T,x)\nabla V)=0$ together with the transient heat-diffusion equation on the 2D device cross-section, with the local Joule heating $Q(x)=\sigma |\nabla V|^2$ serving as the heat source. 

Temperature-dependent silicon properties were included throughout: the thermal conductivity $\kappa(T)$ followed the thin-film phonon model \cite{Liu2006ModelingTemperature}, incorporating layer-thickness and doping corrections, while the electrical conductivity $\sigma (T)$ accounted for both the rise in intrinsic carrier concentration $(n_i)$ \cite{Sze2006PhysicsDevices} and the change in carrier mobility \cite{Masetti1983ModelingSilicon, Canali1975ElectronTemperature} at elevated temperature (see Supplementary Information \textbf{Section S14} for the complete set of parameters). The measured drive voltage was applied as the condition at the TiN/Si contacts, and the substrate and metal contacts were held at a fixed ambient temperature (300 K) as heat sinks.

\medskip
\textbf{Supporting Information} \par 
Supporting Information is available as a separate file accompanying this preprint, or from the corresponding author.

\medskip
\textbf{Conflict of Interest Statement} \par
The authors declare no competing interests.

\medskip
\textbf{Acknowledgments} \par 
This work was supported by the National Science Foundation under award numbers 2337674, 2329087 and 2210168/2210169, as well as AFOSR Young Investigator Award \#FA9550-24-1-0064. This work was also supported by the Natural Sciences and Engineering Research Council of Canada (NSERC) and the Canada Foundation for Innovation (CFI). BJS is supported by the Canada Research Chairs program.

Work performed in the University of Pittsburgh Nanofabrication and Characterization Core Facility (RRID:SCR\_05124) and services and instruments used in this project were graciously supported, in part, by the University of Pittsburgh. The authors wish to acknowledge Dr. Susheng Tan for his helpful assistance in TEM sample preparation and data collection.

Chi-Yi Kao acknowledges support from the National Science and Technology Council (NSTC), Taiwan, under Grant Nos. 114-2221-E-002-216-MY3 and 115-2917-I-002-021. 
\medskip

%

\bibliographystyle{unsrtnat}
\bibliography{references}


\end{document}